# Regime-switching constrained viscosity solutions approach for controlling dam-reservoir systems

Hidekazu Yoshioka[1,*], Yumi Yoshioka[1]

[1] Assistant Professor, Graduate School of Natural Science and Technology, Shimane University, Nishikawatsu-cho 1060, Matsue, 690-8504, Japan
* Corresponding author: yoshih@life.shimane-u.ac.jp

**Abstract**

A new stochastic control problem of a dam-reservoir system installed in a river is analyzed both mathematically and numerically. Water balance dynamics of the reservoir are piece-wise deterministic and are driven by a stochastic regime-switching inflow process. The system is controlled to balance among the operation purpose and the internal and downstream environmental conditions. Finding the optimal operation policy of the system reduces to solving an optimality equation with a discontinuous Hamiltonian, which is a system of nonlinear degenerate parabolic (or hyperbolic) equations. We show that the optimality equation has at most one constrained viscosity solution and find the solution explicitly under certain conditions. The model is applied to numerical computation of the operation policy of an existing dam-reservoir system using a high-order finite difference scheme. The computational results can suggest how the operation policy should be adapted according to the environmental concerns of the river.

# 1. Introduction

Dam-reservoir systems are fundamental in human lives, providing water resources [1], mitigating floods [2], generating hydropower [3], and sometimes serving as recreational places [4]. On the other hand, environmental and ecological impacts of creating dams are significant, triggering massive algae bloom in dam-downstream reaches due to low flow discharge [5], altering geomorphology and sediment supply [6], and critically affecting fish migration [7]. Optimization of the dam-reservoir systems to balance between the operation purpose and river environment has long been a central issue in environmental engineering and related research areas.

The stochastic optimal control based on the dynamic programming principle [8] is one of the most successful methodologies for optimization of dam-reservoir systems in rivers. A variety of optimization problems, ranging from simple ones amenable to detailed mathematical analysis [1, 9-11] to practical ones for operational purposes [12-15] have been considered based on the stochastic control and optimization.

We focus on simple mathematical models, because deeply analyzing simplified problems can often provide useful and non-trivial insights into more complicated and practical cases. Even relatively simple problems of managing single reservoir have rich and non-trivial mathematical structures [16]. Modern mathematical tools related to the stochastic dynamic programming and degenerate parabolic partial differential equations (PDEs), such as stochastic differential equations [17, 18] and viscosity solutions [19], are useful in analyzing these problems. Abbramov [20] formulated a queue model of a large reservoir receiving stochastic inflows and carried out asymptotic analysis of its optimal operation policy. Kharroubi [21] analyzed a general stochastic optimal switching model from the viewpoint of constrained viscosity solutions to a degenerate parabolic quasi-variational inequality with a particular focus on hydropower generation using a reservoir. Jiang et al. [22] considered an exactly-solvable unified game-theoretic control model of mitigating watershed pollution through water diversions.

Recently, Yoshioka and Yoshioka [11] considered a stochastic control model of single dam-reservoir systems balancing the human activities and river water environment. Their model is based on a stochastic differential equation (SDE) governing water balance dynamics in a reservoir. The outflow discharge from the dam is the control variable. The model is exactly-solvable under an ergodic long-time limit under some simplifications. Severe drawbacks of the model and their approach are the following three-holds. Firstly, they analyzed temporally-constant (and thus time-independent) inflows that are less realistic. Secondly, the control set is regularized when the water volume in the reservoir is close to the empty or full, with which the Hamiltonian associated with the optimality equation becomes globally continuous. The continuity plays an indispensable role in the analysis of the optimality equation, such as unique solvability and convergence of numerical solutions in a viscosity sense [23]. The third drawback is the numerical scheme that they utilized. The scheme is a seemingly monotone finite difference scheme easy to implement, while such schemes are sometimes not sufficiently accurate especially against non-smooth solutions. Overcoming these issues would contribute to establishment of more useful and realistic models for optimization of dam-reservoir systems from a mathematical side. This is the motivation of our paper. Note that similar issues can be encountered in optimization of resource storage systems for natural gasses [24] and oils [25].

The objectives as well as contributions of this paper are formulation, analysis, and application of a stochastic control model of a dam-reservoir system receiving stochastic (and thus temporally varying) inflows. The water balance dynamics in the reservoir are piece-wise deterministic subject to a regime-switching inflow process. This system is considered as an SDE of a hybrid type driven by a continuous-time Markov chain [18]. The Markov chain can naturally represent time-dependent inflows [26-28]. The first drawback of the previous model is then improved. A consequence of hybridizing the dynamics is that the resulting optimality equation does not become single PDE like the conventional control problems [17] but a weakly-coupled system of PDEs [29]. The water balance dynamics for small and large water volume are handled physically based on the concept of constrained viscosity solutions [30] without relying on the regularization of the control set [11, 24]. This concept enables us to deal with the cases where the water volume is close to empty or full in a physically reasonable manner. In addition, it is suited to analysis and approximation of the optimality equation in a weak, i.e., a viscosity sense. To the best of the authorsøknowledge, the proposed mathematical model is new.

We also demonstrate that the model can be applied to numerical computation on realistic problems. The degenerate parabolic form of the optimality equation allows us to apply high-resolution numerical schemes to its discretization. A monotone numerical scheme equipped with a weighted-essentially non-oscillatory (WENO) reconstruction [31] turns out to be an effective numerical method for our problem. Similar numerical methods have been proven to efficiently approximate a variety of degenerate parabolic and hyperbolic problems [32-34]. We show that our problem is no exception and the WENO reconstruction indeed improves accuracy of the scheme. The model parameters are identified at an existing dam in Japan and the optimal policy balancing between its operation purpose, and ecosystems inside and downstream of the reservoir is explored numerically. Our contribution thus covers mathematical, numerical, and practical aspects of optimal control of dam-reservoir systems.

The rest of this paper is organized as follows. Our mathematical model is presented in

Section 2. Mathematical analysis of the optimality equation focusing on its exact solutions and unique solvability is carried out in Section 3. The presented model is applied to test and realistic cases in Section 4. Computational performance of the WENO reconstruction is assessed using the exact solution. The computational results of the realistic problem can suggest how the operation policy should be adapted according to the environmental concerns of the river. Summary and future perspective of our research are presented in Section 5.

## 2. Mathematical model

### 2.1 Water balance dynamics

Our problem setting is explained in this sub-section. **Figure 1** is the conceptual diagram of our problem. Let $t \geq 0$ be time. We consider a single dam-reservoir system receiving a time-dependent inflow discharge $\theta_t > 0$, which is a piece-wise constant variable following a continuous-time $(I+1)$-regime Markov chain $\alpha = (\alpha_t)_{t \geq 0}$ with $I \in \mathbb{N}$. A natural filtration generated by $\alpha$, which is the source of stochasticity in our model, is denoted as $\mathcal{F} = (\mathcal{F}_t)_{t \geq 0}$. The regimes $i \in J = \{0,1,2,...,I\}$ are determined as follows: $\alpha_t = i$ if $\theta_t \in R_i = [\Delta_i, \Delta_{i+1})$ ($i \in J$), where $\{\Delta_i\}_{i=0,1,2,...}$ with $\Delta_0 = 0$ and $\Delta_{I+1} = +\infty$ is a strictly increasing sequence. The switching rates from $i$ to $j$ is denoted as $\lambda_{ij} \geq 0$. Here, $\lambda_{ij} \Delta t$ with a small $\Delta t > 0$ asymptotically represents the conditional probability that the regime switching from $i$ to $j$ ($i \neq j$) occurs during the time interval $(t, t+\Delta t)$ provided that the regime is $i$ at the time $t$. Set a representative value $Q_i \in R_i$ of the inflow discharge for each $i \in J$, which can be a midpoint or some average. We then describe the inflow process as $\theta_t = \sum_{i=0}^{I} \chi_{\{\alpha_t = i\}} Q_i$, where $\chi_S$ is the indicator function of the set $S$; $\chi_S = 1$ if $S$ is true while $\chi_S = 0$ otherwise.

The state variable to be observed by the decision-maker, the operator of the dam, is the inflow regime $\alpha_t$ and the water volume $V_t$ of the reservoir at each $t$. The capacity of the reservoir is prescribed as $\bar{V} > 0$. Physically, the range of $V_t$ should be the compact set $\Omega = [0, \bar{V}]$. The outflow discharge is denoted as $q_t$ at $t$. Its possible range is $A = [\underline{q}, \bar{q}]$ with some constants $0 \leq \underline{q} < \bar{q}$ determined from technological constraints of the dam, which has to be modified when the reservoir is at a full or empty as discussed later. There may exist residual elements affecting the water balance dynamics, such as the direct rainfall and evaporation from the water surface. They are represented in a lumped manner as $\varpi(t,V_t)$ at $t$, which is assumed to be Lipschitz continuous with respect to both the first and second arguments [11]. Lipschitz continuous functions can cover very smooth (infinitely continuously differentiable) functions to continuous functions that are close to be discontinuous. We assume $\bar{q} > Q_I + \sup_{t \geq 0, V \in \Omega} \varpi(t,V)$ and $\underline{q} < Q_0 + \inf_{t \geq 0, V \in \Omega} \varpi(t,V)$ meaning that the dam has a satisfactory capacity to handle the inflow process. This assumption may not be true in some applications. We assume this for convenience, but we will encounter problems not satisfying it in future. However, such problems, of course interesting, are beyond the scope of this paper.

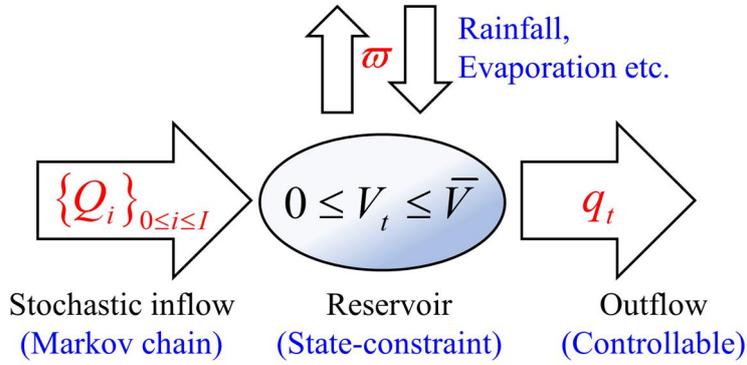

**Figure 1.** Conceptual diagram of the water balance dynamics considered in this paper.

*Remark 1*

In our model application in Section 4, we assume that $\varpi(t,V_t)$ is small and omit it, but the presence of this coefficient does not affect our mathematical and numerical analyses. In addition, it may be possible to aggregate $\varpi$ to the inflow discharge $\theta$.

Considering the water balance in the reservoir leads to the regime-switching SDE

$$dV_t = \left(\theta_t - q_t + \varpi(t,V_t)\right)dt = \left(\sum_{i=0}^{I} \chi_{\{\alpha_t = i\}} Q_i - q_t + \varpi(t,V_t)\right)dt, \quad t > 0 \quad (1)$$

subject to an initial condition $V_0 \in \Omega$ and $\alpha_0 \in J$. This equation simply means that the water balance dynamics are driven by the inflow, outflow, rainfall, and evaporation etc. The admissible set of the outflow discharge $q = (q_t)_{t \geq 0}$ has to be specified to fully characterize the water balance dynamics. As mentioned above, we should consider the operation policy such that the constraint $V_t \in \Omega$ is satisfied a.s. $t \geq 0$. The admissible set of $q$ is denoted as $\mathcal{Q}$, and is defined as

$$\mathcal{Q} = \{q = (q_t)_{t\geq 0} | q \text{ is progressively measurable w.r.t. } \mathcal{F}, V_t \in \Omega \text{ a.s. } t \geq 0.\}. \tag{2}$$

For convenience, we assume that there exists a unique strong solution $V = (V_t)_{t\geq 0}$ for each $q \in \mathcal{Q}$.

The constraint $V_t \in \Omega$ physically requires to modify the range $A$ to depend on $V_t$:

$$A(V) = \begin{cases} [\theta_t + \varpi(t,\bar{V}), \bar{q}] & (V = \bar{V}) \\ [\underline{q}, \bar{q}] & (0 < V < \bar{V}), \ V \in \Omega. \\ [\underline{q}, \theta_t + \varpi(t,0)] & (V = 0) \end{cases} \tag{3}$$

The set $A(V)$ is compact with respect to $V \in \Omega$, but it is not continuous at $V = 0, \bar{V}$ (Here, the notation $V = 0, \bar{V}$ represents $V = 0$ and $V = \bar{V}$). This discontinuity comes from the naïve physical assumption that the water volume in the reservoir should not over- and under-shoot the full and empty states, respectively. In this way, we formally require the inequalities corresponding to the confinement

$$\frac{dV_t}{dt} = \theta_t - q_t + \varpi(t,V_t) \leq 0 \quad \text{if } V_t = \bar{V} \tag{4}$$

and

$$\frac{dV_t}{dt} = \theta_t - q_t + \varpi(t,V_t) \geq 0 \quad \text{if } V_t = 0. \tag{5}$$

A remarkable difference between the previous [11] and present approaches is that the former artificially regularized the set $A(V)$ so that it becomes continuous with respect to $V \in \Omega$, while the latter does not use such a technique. A regularization of the control set has also been used in Shardin and Szölgyenyi [24]. The price to be paid for not using the regularization is that the optimality equation derived later has to be carefully handled at the boundary points. We resolve this issue by utilizing the concept of constrained viscosity solutions. In this way, we do not have to directly handle the discontinuity in (3) at the boundary points.

*Remark 2*

The regime-switching representation of the inflow process can be seen as a semi-discrete counterpart of the continuous-state ones of several theoretical models [35, 36] or numerical ones [37].

**2.2 Objective function and the optimality equation**

The objective function $\phi = \phi(t,v,i;q)$ is a functional of the current observation value

$(V_t, \alpha_t) = (v, i)$ and the control $q \in \mathcal{Q}$. In this paper, $\phi$ is set as the expected sum of the terms on the discharge and water volume:

$$\phi(t,v,i;q) = \mathbb{E}_{t,v,i}\left[\int_t^T e^{-\delta(s-t)} f(s,q_s,\alpha_s)\mathrm{d}s + \int_t^T e^{-\delta(s-t)} g(s,V_s,\alpha_s)\mathrm{d}s \middle| (V_t,\alpha_t)=(v,i)\right], \quad t \leq T, \; v \in \Omega, \; i \in J, \; q \in \mathcal{Q}, \quad (6)$$

where $\delta \geq 0$ is the discount rate and $T > 0$ is a prescribed terminal time that may be $\infty$. In the latter case, we must assume the positivity $\delta > 0$. The first and second terms in the right-hand side of (6) measure the disutilities due to environmental and ecological conditions of the downstream of and inside the reservoir as explained below.

We assume that the first term contains the two terms: $f = f_1 + f_2$. The first one measures the deviation between the target and chosen outflow discharges

$$f_1(s,q_s,\alpha_s) = \frac{1}{m+1}\left|\ddot{q}(s,\alpha_s) - q_s\right|^{m+1} \quad (7)$$

with a constant $m > 0$ and the prescribed target discharge $\ddot{q}(s,\alpha_s) > 0$ depending on the inflow discharge and thus on the Markov chain $\alpha$. In some real cases, $\ddot{q}(s,\alpha_s) = \theta_s$ except for extremely large inflows due to floods [5, 38]. For reservoirs aiming at supplying water resources, the target discharge can be determined according to the demand by the stakeholders [1, 39]. The second one concerns with the outflow discharge smaller than a threshold, below which the downstream environmental conditions may be severely affected:

$$f_2(s,q_s,\alpha_s) = \frac{w}{n+1}\max\{\tilde{q}(s,\alpha_s) - q_s, 0\}^{n+1} \quad (8)$$

with a constant $n > 0$, a weight constant $w > 0$, and the prescribed threshold discharge $\tilde{q}(s,\alpha_s) > 0$. This threshold can be determined according to the objective of the dam operation. As an example of nuisance benthic algae bloom in dam-downstream reaches, there exist some threshold discharge above which the growth rate of the algae becomes negative [5, 38]. Therefore, the threshold discharge $\tilde{q}_s$ can be determined according to the target aquatic species living in dam-downstream reaches. The threshold discharge can also be determined considering the minimal flow discharge enforced by a law [40, 41].

The second term $g$ is set as a penalty incurred when the water volume $V_t$ is not in a prescribed range $\omega_t = [a_t, b_t]$ with continuous and smooth time-varying parameters $0 < a_t < b_t < \overline{V}$. The range $\omega_t$ can be determined by the operation purpose of the system. A too large water volume may physically damage the dam [20] and a too small water volume may threaten aquatic species in the reservoir [42]. In addition, ecologically friendly reservoir operations are preferred if keystone aquatic species are spawning in the reservoir [43]. We set the penalization as

$$g(s, V_s, \alpha_s) = y \chi_{\{V_s \notin \omega_s\}} \qquad (9)$$

with a prescribed weight constant $y > 0$. This choice of the penalization is simple and makes the model be analytically tractable as shown in the next section, but it emerges as a discontinuity in the Hamiltonian of our optimality equation.

The minimized performance index $\phi$ with respect to $q \in \mathcal{Q}$ is called the value function:

$$\Phi(t, v, i) = \inf_{q \in \mathcal{Q}} \phi(t, v, i; q), \quad t \leq T, \ v \in \Omega, \ i \in J. \qquad (10)$$

The goal of the presented optimization problem is to find the optimal control, which is denoted as $q^* \in \mathcal{Q}$, to achieve the minimization in (10). Set $D = [0, T) \times \Omega \times J$ and the Hamiltonian $H: D \times \mathbb{R} \times \mathbb{R}^I \times \mathbb{R} \to \mathbb{R}$ as

$$H\left(t, v, i, \varphi, \{\psi_j\}_{0 \leq j \leq I, j \neq i}, p\right) = \delta \varphi + \sum_{0 \leq j \leq I, j \neq i} \lambda_{ij}(\varphi - \psi_j) \\ - \min_{q \in A(v)} \{(Q_i - q + \varpi(t, v))p + f(t, q, i) + g(t, v, i)\} \qquad (11)$$

The dynamic programming principle formally leads to the optimality equation governing $\Phi$:

$$-\frac{\partial \Phi_i}{\partial t} + H\left(t, v, i, \Phi_i, \{\Phi_j\}_{0 \leq j \leq I, j \neq i}, \frac{\partial \Phi_i}{\partial v}\right) = 0 \ \text{ in } \ D \qquad (12)$$

subject to the terminal condition

$$\Phi_i = 0 \ \text{ for } \ \{t = T\} \times \Omega \times J, \qquad (13)$$

where $\Phi_i = \Phi(t, v, i)$.

The optimality equation (12) is a system of weakly-coupled nonlinear degenerate parabolic (or hyperbolic) PDEs. Its solutions are expected not to be sufficiently smooth such that they satisfy (12) point-wise. In general, degenerate parabolic PDEs admit only non-smooth solutions in a viscosity sense [19]. In the next section, we analyze the optimality equation (12) in a viscosity sense. Notice that the Hamiltonian $H$ is discontinuous at $v = a, b$ by (9).

The optimal outflow discharge $q^* = q^*(t, v, i)$ as a function of the time $t$, water volume $v$, regime $i \in J$ is formally obtained as

$$q^*(t, v, i) = \arg\min_{q \in A(v)} \left\{(Q_i - q + \varpi(t, v))\frac{\partial \Phi_i}{\partial v} + f(t, q, i)\right\} = \arg\min_{q \in A(v)} \left\{-q\frac{\partial \Phi_i}{\partial v} + f(t, q, i)\right\}. \quad (14)$$

In this sense, the optimal outflow discharge $q^*$ is found as a quantity based on the value function $\{\Phi_i\}_{0\leq i\leq I}$. In Section 4, we present an explicit algorithm to compute $\{\Phi_i\}_{0\leq i\leq I}$ and $q^*$.

*Remark 3*

Each term $\lambda_{ij}(\Phi_i - \Phi_j)$ in the optimality equation (11) can be seen as a discrete counterpart of partial differential terms on some numerical grid [44, 45] of the inflow discharge. This is in accordance with **Remark 2**, showing a hybrid nature of our model.

A key inequality on the Hamiltonian $H$ is presented, which plays an essential role in the proof of comparison argument of the optimality equation.

*Lemma 1*

*There exists a constant $H_C > 0$ such that*

$$\left| H\left(t,u,i,\varphi,\{\psi_j\}_{0\leq j\leq I, j\neq i}, p\right) - H\left(t,v,i,\varphi,\{\psi_j\}_{0\leq j\leq I, j\neq i}, p\right) \right| \leq H_C |u-v||p| \quad (15)$$

*for all $t \in [0,T]$, $u,v \in (0,a)$, $p,\varphi,\psi_j \in \mathbb{R}$, and $i,j \in J$. The same statement holds true with $u,v \in (a,b)$ and $u,v \in (b,\overline{V})$.*

**(Proof of Lemma 1)**

Assume $u,v \in (0,a)$. The proofs for the cases $u,v \in (a,b)$ and $u,v \in (b,\overline{V})$ are essentially the same. Firstly, $A(u) = A(v)$ and $g(t,u,i) = g(t,v,i)$ by $u,v \in (0,a)$. We get

$$\begin{aligned}
&\left| H\left(t,u,i,\varphi,\{\psi_j\}_{0\leq j\leq I, j\neq i}, p\right) - H\left(t,v,i,\varphi,\{\psi_j\}_{0\leq j\leq I, j\neq i}, p\right) \right| \\
&= \left| \min_{q\in A(u)}\{(Q_i - q + \varpi(t,u))p + f(t,q,i)\} - \min_{q\in A(u)}\{(Q_i - q + \varpi(t,v))p + f(t,q,i)\} \right| \\
&= \left| (\varpi(t,u) - \varpi(t,v))p + \min_{q\in A(u)}\{-qp + f(t,q,i)\} - \min_{q\in A(u)}\{-qp + f(t,q,i)\} \right| \quad (16)\\
&= |\varpi(t,u) - \varpi(t,v)||p| \\
&\leq C_1 |u-v||p|
\end{aligned}$$

because of the Lipschitz continuity of $\varpi$. We can set $H_C = C_1$.

*Remark 4*

We focus on the optimality equation but not the underlying dynamic programming principle. This is rather standard for cases where $f$ is continuous, but may not be so if $f$ is discontinuous. The latter case can be handled in the framework of the optimal controls with discontinuous coefficients [46].

## 3. Mathematical analysis

### 3.1 Constrained viscosity solutions

In this section, we analyze the optimality equation (12) from a viscosity viewpoint under state constraint [30]. In our case, the constraint is $V_t \in \Omega$. The following definition of viscosity solutions is the starting point of the mathematical analysis. A key in the definition is to asymmetrically define viscosity sub-solutions and super-solutions, respectively. The super-solution property is required over the domain $\Omega$, while the sub-solution property is not required along the boundaries $v = 0, \overline{V}$ of $\Omega$. Formally, the optimality equation (12) is replaced by a one-sided inequality along the boundaries $v = 0, \overline{V}$. Set $\mathring{\Omega} = (0, \overline{V})$. In what follows, $USC(\cdot)$ and $LSC(\cdot)$ are collections of the upper-semicontinuous and lower-semicontinuous functions.

*Definition 1*

*A set of functions $\Psi = \{\Psi_i\}_{0 \le i \le I}$ with $\Psi_i \in USC([0,T] \times \Omega)$ and $\Psi_i \le 0$ for $t = T$ is called a viscosity sub-solution if for all $(t_0, v_0, i_0) \in [0,T) \times \mathring{\Omega} \times J$ and for all $\varphi = \{\varphi_i\}_{0 \le i \le I}$ with $\varphi_i \in C^1([0,T] \times \Omega)$, $\varphi_i - \Psi_i$ is locally minimized at $(t,v) = (t_0, v_0)$ as $\varphi_{i_0}(t_0, v_0) - \Psi_{i_0}(t_0, v_0) = 0$ and*

$$-\frac{\partial \varphi_{i_0}}{\partial t} + H_*\left(t_0, v_0, i_0, \varphi_{i_0}(t_0, v_0), \{\varphi_j(t_0, v_0)\}_{0 \le j \le I, j \ne i_0}, \frac{\partial \varphi_{i_0}}{\partial v}(t_0, v_0)\right) \le 0. \qquad (17)$$

*A set of functions $\Psi = \{\Psi_i\}_{0 \le i \le I}$ with $\Psi_i \in LSC([0,T] \times \Omega)$ and $\Psi_i \ge 0$ for $t = T$ is called a viscosity super-solution if for all $(t_0, v_0, i_0) \in [0,T) \times \Omega \times J$ and for all $\varphi = \{\varphi_i\}_{0 \le i \le I}$ with $\varphi_i \in C^1([0,T] \times \Omega)$, $\varphi_i - \Psi_i$ is locally maximized at $(t,v) = (t_0, v_0)$ as $\varphi_{i_0}(t_0, v_0) - \Psi_{i_0}(t_0, v_0) = 0$ and*

$$-\frac{\partial \varphi_{i_0}}{\partial t} + H^*\left(t_0, v_0, i_0, \varphi_{i_0}(t_0, v_0), \{\varphi_j(t_0, v_0)\}_{0 \le j \le I, j \ne i_0}, \frac{\partial \varphi_{i_0}}{\partial v}(t_0, v_0)\right) \ge 0. \qquad (18)$$

*A function* $\Psi \in C([0,T] \times \Omega)$ *is a viscosity solution if it is a viscosity sub-solution as well as a viscosity super-solution.*

Notice that the sub-solution property is not required at $v = 0, \bar{V}$. This relaxed definition comes from the constraint of the control set to confine the state variable in the domain [30].

For later use, we also give the steady counterpart of the definition, which corresponds to the problem with time-independent coefficients and $T = +\infty$. The dependence of the quantities on $t$ is effectively omitted.

## *Definition 2*

*A set of functions* $\Psi = \{\Psi_i\}_{0 \leq i \leq I}$ *with* $\Psi_i \in USC(\Omega)$ *is called a viscosity sub-solution if for all* $(v_0, i_0) \in \ddot{\Omega} \times J$ *and for all* $\varphi = \{\varphi_i\}_{0 \leq i \leq I}$ *with* $\varphi_i \in C^1(\Omega)$, $\varphi_i - \Psi_i$ *is locally minimized at* $v = v_0$ *as* $\varphi_{i_0}(v_0) - \Psi_{i_0}(v_0) = 0$ *and*

$$H_*\left(v_0, i_0, \varphi_{i_0}(v_0), \{\varphi_j(v_0)\}_{0 \leq j \leq I, j \neq i_0}, \frac{\partial \varphi_{i_0}}{\partial v}(v_0)\right) \leq 0. \quad (19)$$

*A set of functions* $\Psi = \{\Psi_i\}_{0 \leq i \leq I}$ *with* $\Psi_i \in LSC(\Omega)$ *is called a viscosity super-solution if for all* $(v_0, i_0) \in \Omega \times J$ *and for all* $\varphi = \{\varphi_i\}_{0 \leq i \leq I}$ *with* $\varphi_i \in C^1(\Omega)$, $\varphi_i - \Psi_i$ *is locally maximized at* $v = v_0$ *as* $\varphi_{i_0}(v_0) - \Psi_{i_0}(v_0) = 0$ *and*

$$H^*\left(v_0, i_0, \varphi_{i_0}(v_0), \{\varphi_j(v_0)\}_{0 \leq j \leq I, j \neq i_0}, \frac{\partial \varphi_{i_0}}{\partial v}(v_0)\right) \geq 0. \quad (20)$$

*A set of functions* $\Psi = \{\Psi_i\}_{0 \leq i \leq I}$ *with* $\Psi_i \in C(\Omega)$ *is a viscosity solution if it is a viscosity sub-solution as well as a viscosity super-solution.*

## 3.2 Exact solution

For $\delta > 0$, we present an explicit constrained viscosity solution to the steady problem

$$H\left(v, i, \Phi_i, \{\Phi_j\}_{0 \leq j \leq I}, \frac{\partial \Phi_i}{\partial v}\right) = 0 \quad \text{in } \Omega \times J \quad (21)$$

in the sense of **Definition 2**. Exact solutions to optimality equations in stochastic control problems are often derived under simplified conditions, but can give useful insights into properties of the equations [47, 48]. Such solutions cam be used in verifying computational performance of numerical schemes as well, as demonstrated in this paper.

Set $\omega = [a,b]$ with $0 < a < b < \bar{V}$, $m = n = 1$, $\varpi \equiv 0$, and $\ddot{q} = \tilde{q} = Q_i$ for each regime $R_i$. This parameter setting corresponds to the problem where the water balance dynamics are dominated by the inflow and outflow, and the target and thresholds discharges equal. Therefore, the outflow discharge smaller than the threshold (and target) discharge is more penalized than that larger than the threshold (and target) discharge. In this way, the disutility potentially caused by the outflow discharge is asymmetric. This is somewhat an artificial setting because the threshold discharge should depend on the biological and physical parameters on the algae, but not on the inflow regimes. Nevertheless, the obtained solution is non-trivial and demonstrates a state-dependent optimal control $q^*$.

We show that the steady optimality equation (21) admits the following exact constrained viscosity solution. Notice that it is continuous but only piece-wise smooth, meaning that it is not a classical smooth solution.

*Proposition 1*

*Assume*

$$\frac{1+w}{m^m(m+1)} \delta^{m+1} \max\{a, \bar{V}-b\}^{m+1} \leq y \leq \frac{m}{(m+1)(1+w)^{\frac{1}{m}}} \min_{i \in J}\{Q_i, \bar{q}-Q_i\}^{m+1}. \quad (22)$$

*Then, the following set of functions $\{\ddot{\Phi}_i(v)\}_{0 \leq i \leq I}$ defined in $\Omega$ is a constrained viscosity solution to the steady optimality equation (21): $\ddot{\Phi}_i = \ddot{\Phi}(v)$, where*

$$\ddot{\Phi}(v) = \begin{cases} f_L(v) & (0 \leq v < a) \\ 0 & (a \leq v \leq b) \\ f_R(v) & (b < v \leq \bar{V}) \end{cases} \quad (23)$$

*with*

$$f_L(v) = \frac{y}{\delta} - \left[\left(\frac{y}{\delta}\right)^{\frac{1}{m+1}} - C(a-v)\right]^{m+1}, \quad f_R(v) = \frac{y}{\delta} - \left[\left(\frac{y}{\delta}\right)^{\frac{1}{m+1}} - C(v-b)\right]^{m+1}, \quad (24)$$

*and a constant $C = \frac{(1+w)^{\frac{1}{m+1}}}{m+1}\left(\frac{m+1}{m}\delta\right)^{\frac{m}{m+1}}$.*

**(Proof of Proposition 1)**

By (23), we see $\ddot{\Phi} \in C(\Omega)$ and is continuously differentiable except at $v = a, b$. In addition, we have $\ddot{\Phi}_i - \ddot{\Phi}_j = 0$ $(i, j \in J)$ and thus all the terms of the form $\lambda_{i,j}(\ddot{\Phi}_i - \ddot{\Phi}_j)$ vanish from the equation. It is straightforward to check that $\{\ddot{\Phi}_i(v)\}_{0\leq i\leq I}$ satisfies (21) in the classical sense except at $v=a,b$. The viscosity super-solution property is directly verified with the help of the shape of $\chi_{\{v\notin[a,b]\}}$. At the kink points $v=a,b$, there exists no test function for viscosity sub-solutions, and thus the sub-solution property is trivial.

□

The condition (22) means that the penalization on the water volume is not so large ($y$ is moderately small) and the decision-maker controls the dam from a sufficiently long-term viewpoint ($\delta$ is sufficiently small). Therefore, the penalization should be neither too strong nor too small to justify this exact solution. The exact solution in **Proposition 3.1** corresponds to the case where the value function is independent from the regimes. On the other hand, the optimal outflow discharge $q^*$ computed from the solution is different among the regimes. Indeed, we can substitute $\ddot{\Phi}$ into (14) and obtain

$$q_i^*(v) = Q_i + \begin{cases} g_L(v) & (0 \leq v < a) \\ 0 & (a \leq v \leq b) \\ g_R(v) & (b < v \leq \overline{V}) \end{cases} \quad (25)$$

with

$$g_L(v) = -\left((m+1)C\right)^{\frac{1}{m}}\left[\left(\frac{y}{\delta}\right)^{\frac{1}{m+1}} - C(a-v)\right], \quad g_R(v) = \left((m+1)C\right)^{\frac{1}{m}}\left[\left(\frac{y}{\delta}\right)^{\frac{1}{m+1}} - C(v-b)\right]. \quad (26)$$

**Figure 2** plots the exact solution and the corresponding optimal discharge. The parameter values are as follows: $\overline{V}=1$, $(a,b)=(0.3,0.7)$, $m=n=1$, $y=0.5$, $\underline{q}=0$, $\overline{q}=4$, $\tilde{q}=\ddot{q}=Q_0=1$, $w=0.4$, and $\delta=0.1$. It should be noted that the exact solution and its derivation constraints do not depend on the switching rates $\lambda_{i,j}$.

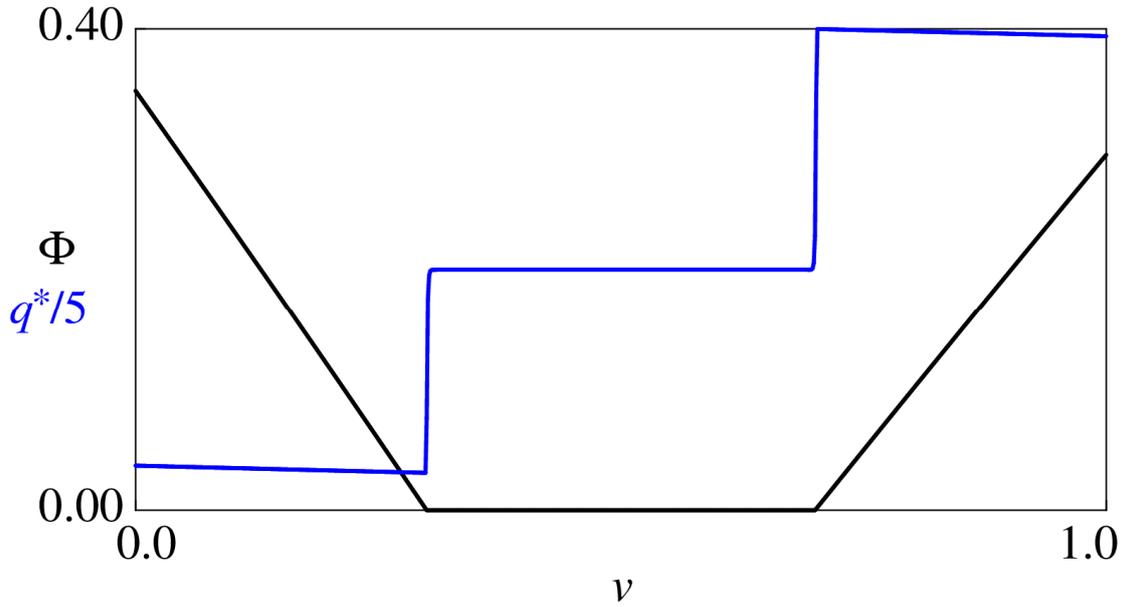

**Figure 2.** The exact viscosity solution $\Phi$ (Black) and the corresponding optimal control $q^*$ ($i=0$, Blue). The lateral axis is the water volume and the vertical axis represents the solution and optimal control.

### 3.3 Comparison theorem

We present a comparison theorem of the steady optimality equation (21). The proof for the original time-dependent optimality equation (12) based on **Definition 1** is essentially the same because our terminal condition is standard, and is not presented here [19].

The following **Proposition 2** states that there exists at most one constrained viscosity solutions to the steady optimality equation (21). Its proof is based on the standard technique of doubling the variables [19]. However, we need to consider the interior discontinuity of $H$ and the asymmetric definitions between the sub- and super-solutions **(Definition 2)**. These two difficulties are resolved through employing the following techniques. The discontinuity is handled with a specialized auxiliary function for the doubling the variables technique (Proof of Theorem 11.4 of Calder [49]). The asymmetry of the definitions is handled with another specialized auxiliary function (Proof of Theorem 2.2 of Katsoulakis [30]).

*Proposition 2*

*For any viscosity sub-solution $\{\varphi_i\}_{0 \leq i \leq I}$ and viscosity super-solution $\{\psi_i\}_{0 \leq i \leq I}$, $\varphi_i \leq \psi_i$ ($v \in \Omega$, $0 \leq i \leq I$).*

**(Proof of Proposition 2)**

The proof uses a contradiction argument. Assume that there is some $(v_0, i_0) \in \Omega \times J$ such that $\varphi_{i_0}(v_0) - \psi_{i_0}(v_0) = \gamma > 0$. Without any loss of generality, set $\gamma = \max_{(v,i) \in \Omega \times J}(\varphi_i(v) - \psi_i(v))$. We separately consider the exclusive cases: (a) $v_0 = a$, (b) $v_0 = b$, (c) $v_0 = 0$, (d) $v_0 = \overline{V}$, and (e) Otherwise.

The case (e) is the simplest case. Set the auxiliary function $f_\alpha : \Omega \times \Omega \to \mathbb{R}$ as follows:

$$f_\alpha(u,v) = \varphi_{i_0}(u) - \psi_{i_0}(v) - \frac{\alpha}{2}|u-v|^2 \qquad (27)$$

with some $\alpha > 0$. A maximizer of $f_\alpha$ is denoted as $(u_\alpha, v_\alpha)$, which certainly exists because $\Omega$ is compact and $J$ is finite. As in the standard methodology of the comparison [19], we have $(u_\alpha, v_\alpha) \to (v_0, v_0)$ and

$$\frac{\alpha}{2}|u_\alpha - v_\alpha|^2 \to 0 \qquad (28)$$

as $\alpha \to +\infty$. We see that $\varphi_{i_0}(u) - \psi_{i_0}(v_\alpha) - \frac{\alpha}{2}|u - v_\alpha|^2$ is maximized at $u = u_\alpha$ while $-\left(\varphi_{i_0}(u_\alpha) - \psi_{i_0}(v) - \frac{\alpha}{2}|u_\alpha - v|^2\right)$ is minimized at $v = v_\alpha$. Therefore, we can use $\varphi_{i_0}(u) + \frac{\alpha}{2}|u_\alpha - v_\alpha|^2 - \frac{\alpha}{2}|u - v_\alpha|^2$ and $\psi_{i_0}(v) - \frac{\alpha}{2}|u_\alpha - v_\alpha|^2 + \frac{\alpha}{2}|u_\alpha - v|^2$ as test functions of the viscosity sub-solution and the super-solution, respectively. Furthermore, for sufficiently large $\alpha$, we can assume that $v = u_\alpha, v_\alpha$ are different from $v = 0, a, b, \overline{V}$ and that either $u_\alpha, v_\alpha \in (0, a)$, $u_\alpha, v_\alpha \in (a, b)$, or $u_\alpha, v_\alpha \in (b, \overline{V})$. Therefore, $A(u_\alpha) = A(v_\alpha)$ and $g(u_\alpha, i_0) = g(v_\alpha, i_0)$. Then, we get

$$H\left(u_\alpha, i_0, \varphi_{i_0}(u_\alpha), \{\varphi_j(u_\alpha)\}_{0 \leq j \leq I, j \neq i_0}, p_\alpha\right) \leq 0 \qquad (29)$$

and

$$H\left(v_\alpha, i_0, \psi_{i_0}(v_\alpha), \{\psi_j(v_\alpha)\}_{0 \leq j \leq I, j \neq i_0}, p_\alpha\right) \geq 0 \qquad (30)$$

with $p_\alpha = \alpha(u_\alpha - v_\alpha)$. Combining (29) and (30) yields

$$H\left(u_\alpha, i_0, \varphi_{i_0}(u_\alpha), \{\varphi_j(u_\alpha)\}_{0 \leq j \leq I, j \neq i_0}, p_\alpha\right) - H\left(v_\alpha, i_0, \psi_{i_0}(v_\alpha), \{\psi_j(v_\alpha)\}_{0 \leq j \leq I, j \neq i_0}, p_\alpha\right) \leq 0. \qquad (31)$$

Owing to **Lemma 1**, the left-hand side of (31) is calculated as

$$H\left(u_\alpha, i_0, \varphi_{i_0}(u_\alpha), \{\varphi_j(u_\alpha)\}_{0\leq j\leq I, j\neq i_0}, p_\alpha\right) - H\left(v_\alpha, i_0, \psi_{i_0}(v_\alpha), \{\psi_j(v_\alpha)\}_{0\leq j\leq I, j\neq i_0}, p_\alpha\right)$$

$$= \delta\varphi_{i_0}(u_\alpha) + \sum_{0\leq j\leq I, j\neq i_0} \lambda_{i_0 j}\left(\varphi_{i_0}(v_0) - \varphi_j(v_0)\right)$$

$$- \min_{q\in A(u_\alpha)}\left\{\left(Q_{i_0} - q + \varpi(u_\alpha)\right)p_\alpha + f(q, i_0) + g(u_\alpha, i_0)\right\}$$

$$- \left[\begin{array}{l} \delta\psi_{i_0}(v_\alpha) + \sum_{0\leq j\leq I, j\neq i_0} \lambda_{i_0 j}\left(\psi_{i_0}(v_\alpha) - \psi_j(v_\alpha)\right) \\ - \min_{q\in A(v_\alpha)}\left\{\left(Q_{i_0} - q + \varpi(v_\alpha)\right)p_\alpha + f(q, i_0) + g(v_\alpha, i_0)\right\} \end{array}\right]$$

$$= \delta\varphi_{i_0}(u_\alpha) + \sum_{0\leq j\leq I, j\neq i_0} \lambda_{i_0 j}\left(\varphi_{i_0}(v_0) - \varphi_j(v_0)\right) - \left[\delta\psi_{i_0}(v_\alpha) + \sum_{0\leq j\leq I, j\neq i_0} \lambda_{i_0 j}\left(\psi_{i_0}(v_\alpha) - \psi_j(v_\alpha)\right)\right]$$

$$- \left[\min_{q\in A(u_\alpha)}\left\{\left(-q + \varpi(u_\alpha)\right)p_\alpha + f(q, i_0)\right\} - \min_{q\in A(v_\alpha)}\left\{\left(-q + \varpi(v_\alpha)\right)p_\alpha + f(q, i_0)\right\}\right]$$

$$\geq \delta\left(\varphi_{i_0}(u_\alpha) - \psi_{i_0}(v_\alpha)\right) + \sum_{0\leq j\leq I, j\neq i_0} \lambda_{i_0 j}\left(\varphi_{i_0}(v_0) - \varphi_j(v_0)\right) - \sum_{0\leq j\leq I, j\neq i_0} \lambda_{i_0 j}\left(\psi_{i_0}(v_\alpha) - \psi_j(v_\alpha)\right)$$

$$-H_C |u_\alpha - v_\alpha||p_\alpha| \tag{32}$$

By **Lemma 1** and (28), taking the limit $\alpha \to +\infty$ in (32) yields the inequality

$$\delta\gamma + \sum_{0\leq j\leq I, j\neq i_0} \lambda_{i_0 j}\left(\varphi_{i_0}(v_0) - \varphi_j(v_0)\right) - \sum_{0\leq j\leq I, j\neq i_0} \lambda_{i_0 j}\left(\psi_{i_0}(v_0) - \psi_j(v_0)\right) \leq 0. \tag{33}$$

Rearranging (33) yields

$$\delta\gamma + \sum_{0\leq j\leq I, j\neq i_0} \lambda_{i_0 j}\left[\varphi_{i_0}(v_0) - \psi_{i_0}(v_0) - \left(\varphi_j(v_0) - \psi_j(v_0)\right)\right] \leq 0. \tag{34}$$

Now, we have

$$\gamma = \varphi_{i_0}(v_0) - \psi_{i_0}(v_0) \geq \varphi_j(v_0) - \psi_j(v_0), \quad j\in J. \tag{35}$$

Substituting (35) into (34) gives the contradiction $\gamma \leq 0$ by $\delta > 0$. The proof for case (e) is completed.

For the case (a), following Proof of Theorem 11.4 of Calder [49], we set the auxiliary function

$$f_\alpha(u,v) = \varphi_{i_0}(u) - \psi_{i_0}(v) - \frac{\alpha}{2}\left|u - v - \frac{1}{\sqrt{\alpha}}\right|^2 - |u - a|^2. \tag{36}$$

As in case (e), we get

$$\frac{\alpha}{2}\left|u_\alpha - v_\alpha - \frac{1}{\sqrt{\alpha}}\right|^2 + |u_\alpha - a|^2 \to 0 \tag{37}$$

as $\alpha \to +\infty$. Set $p_\alpha = \alpha\left(u_\alpha - v_\alpha - \frac{1}{\sqrt{\alpha}}\right)$. In the present case, we get $u_\alpha > v_\alpha$ for sufficiently large $\alpha$ by (37). Finally, we can get the contradiction $\gamma \leq 0$ using inequalities analogous to (29) and (30). The case (b) can be handled in the same way using the auxiliary function

$$f_\alpha(u,v) = \varphi_{i_0}(u) - \psi_{i_0}(v) - \frac{\alpha}{2}\left|u - v + \frac{1}{\sqrt{\alpha}}\right|^2 - |u-b|^2. \tag{38}$$

The proof of case (c) is a direct application of the Proof of Theorem 2.2 of Katsoulakis [30]. We use the auxiliary function

$$f_\alpha(u,v) = \varphi_{i_0}(u) - \psi_{i_0}(v) - \left|\frac{\varsigma_m}{\varsigma_l}(u-v) - \varsigma_l\right|^2 - \varsigma_l|u-0|^2 \tag{39}$$

with some $m, l \in \mathbb{N}$, where $\{\varsigma_l\}_{l \in \mathbb{N}}$ is a positive decreasing sequence with $\varsigma_l \to 0$ as $l \to +\infty$. Then, we can just follow the proof by Katsoulakis [30] with the help of the inequality (35). A key point is that $u_\alpha > 0$ for each given $m, l \in \mathbb{N}$, meaning that we do not have to handle the sub-solution property on the boundary. Notice that the domain $\Omega$ is simply a 1-D interval and we are considering continuous viscosity solutions, meaning that the assumptions of Proof of Theorem 2.2 of Katsoulakis [30] are satisfied. The case (d) is proven in essentially the same with the case (c) using the auxiliary function

$$f_\alpha(u,v) = \varphi_{i_0}(u) - \psi_{i_0}(v) - \left|\frac{\varsigma_m}{\varsigma_l}(u-v) + \varsigma_l\right|^2 - \varsigma_l|u - \overline{V}|^2. \tag{40}$$

An immediate consequence of **Propositions 1** and **2** is the following result on the function $\ddot{\Phi}$.

*Proposition 3*

*Assume (22). Then, $\{\ddot{\Phi}_i\}_{0 \leq i \leq I}$ is the unique constrained viscosity solution to the steady optimality equation (21).*

## 4. Numerical computation
### 4.1 Discretization
The local Lax-Friedrichs scheme is the simplest numerical scheme for degenerate parabolic and hyperbolic problems. It is monotone, stable, consistent, and thus convergent in the viscosity sense [50]. However, the scheme is too diffusive when computing solutions having sharp and non-smooth profiles like the exact solution $\ddot{\Phi}$ derive above. Enhancing the scheme through an application of the WENO reconstruction possibly realizes a more accurate scheme in return for the loss of monotonicity, which is an indispensable property to prove convergence of numerical solutions in the viscosity sense [31]. The exact solution, which is a steady solution, can be obtained by temporally integrating the discretized system in a sufficiently long time with a sufficiently small time increment if we use the local Lax-Fredric scheme owing to its monotonicity and stability [51]. The enhanced scheme with a WENO reconstruction is not provably convergent partly due to its complexity. We thus experimentally examine its convergence against the exact solution derive earlier. Notice that there exist several mathematical results on convergence of non-monotone schemes, although they do not cover the WENO reconstruction [52, 53].

The discretization of the optimality equation (12) is explained as follows. The scheme we use is the WENO3 reconstruction with the local Lax-Friedrichs finite difference scheme [31], which has been known to perform the third-order spatial accuracy for sufficiently smooth solutions [33, 34]. The domain $\Omega$ is divided into $K+1$ vertices and $K$ cells with $K \geq 4$. The cell length is $\Delta v = \bar{V}/K$. Set $v_k = k\Delta v$ ($0 \leq k \leq K$). Similarly, set the time steps as $\Delta t = T/L$ with $L \in \mathbb{N}$ and $t_l = l\Delta t$ ($0 \leq l \leq L$). The quantity evaluated at $(v_k, t_l)$ is represented using the super-script like $\Phi_i^{(k,l)}$.

We present the discretization for some $i \in J$ because it is essentially the same for the other regimes. Firstly, we explain the discretization for $0 \leq l \leq L-1$ and $2 \leq k \leq K-2$. We use a fully-explicit discretization in time from $t_L = T$ to $t_0 = 0$. For the sake of brevity, set the following quantities based on finite differences:

$$\Delta^+ \Phi_i^{(k,l+1)} = \Phi_i^{(k+1,l+1)} - \Phi_i^{(k,l+1)}, \quad \Delta^- \Phi_i^{(k,l+1)} = \Phi_i^{(k,l+1)} - \Phi_i^{(k-1,l+1)}, \tag{41}$$

$$\Delta^- \Delta^+ \Phi_i^{(k,l+1)} = \Phi_i^{(k+1,l+1)} - 2\Phi_i^{(k,l+1)} + \Phi_i^{(k-1,l+1)}, \tag{42}$$

$$\omega_- = \frac{1}{1+2r_-^2}, \quad r_- = \frac{\varepsilon + \left(\Delta^- \Delta^+ \Phi_i^{(k-1,l+1)}\right)^2}{\varepsilon + \left(\Delta^- \Delta^+ \Phi_i^{(k,l+1)}\right)^2}, \tag{43}$$

$$\omega_+ = \frac{1}{1+2r_+^2}, \quad r_+ = \frac{\varepsilon + \left(\Delta^- \Delta^+ \Phi_i^{(k+1,l+1)}\right)^2}{\varepsilon + \left(\Delta^- \Delta^+ \Phi_i^{(k,l+1)}\right)^2}, \tag{44}$$

$$p_l^{-,(k,l+1)} = \frac{1}{2}\left(\frac{\Delta^+ \Phi_i^{(k-1,l+1)}}{\Delta v} + \frac{\Delta^+ \Phi_i^{(k,l+1)}}{\Delta v}\right) - \frac{\omega_-}{2}\left(\frac{\Delta^+ \Phi_i^{(k-2,l+1)}}{\Delta v} - 2\frac{\Delta^+ \Phi_i^{(k-1,l+1)}}{\Delta v} + \frac{\Delta^+ \Phi_i^{(k,l+1)}}{\Delta v}\right), \tag{45}$$

and

$$p_l^{+,(k,l+1)} = \frac{1}{2}\left(\frac{\Delta^+\Phi_i^{(k-1,l+1)}}{\Delta v} + \frac{\Delta^+\Phi_i^{(k,l+1)}}{\Delta v}\right) + \frac{\varpi_+}{2}\left(\frac{\Delta^+\Phi_i^{(k+1,l+1)}}{\Delta v} - 2\frac{\Delta^+\Phi_i^{(k,l+1)}}{\Delta v} + \frac{\Delta^+\Phi_i^{(k-1,l+1)}}{\Delta v}\right) \quad (46)$$

with the small constant $\varepsilon = 10^{-12}$ to avoid division by zero.

Based on the local Lax-Friedrichs finite difference scheme, the optimality equation (12) at $(v_k, t_l)$ is discretized as

$$-\frac{\Phi_i^{(k,l+1)} - \Phi_i^{(k,l)}}{\Delta t} + H\left(v_k, i, \Phi_i^{(k,l+1)}, \{\Phi_j^{(k,l+1)}\}_{0\le j\le I, j\ne i}, \frac{p_l^{+,(k,l+1)} + p_l^{-,(k,l+1)}}{2}\right)$$
$$+ \frac{D_i^{(k,l)}}{2}\left(p_l^{+,(k,l+1)} - p_l^{-,(k,l+1)}\right) = 0 \quad (47)$$

where $D_i^{(k,l)}$ is the numerical viscosity coefficient defined by

$$D_i^{(k,l)} = \max_{p\in[p_{\min},p_{\max}]}\left|\frac{\partial H}{\partial p}\left(v_k, i, \Phi_i^{(k,l+1)}, \{\Phi_j^{(k,l+1)}\}_{0\le j\le I, j\ne i}, p\right)\right| \quad (48)$$

with the WENO-reconstructed variables

$$p_{\min} = \min\{p_l^{+,(k,l+1)}, p_l^{-,(k,l+1)}\} \quad \text{and} \quad p_{\max} = \max\{p_l^{+,(k,l+1)}, p_l^{-,(k,l+1)}\}. \quad (49)$$

The point value $q^*(t_l, v_k, i,)$ of the optimal outflow discharge $q^*$ is computed as

$$q^*(t_l, v_k, i,) = \arg\min_{q\in A(v_k)}\left\{(Q_i - q + \varpi(t_l, v_k))\frac{p_l^{+,(k,l+1)} + p_l^{-,(k,l+1)}}{2} + f(t_l, q, i)\right\}$$
$$= \arg\min_{q\in A(v_k)}\left\{-q\frac{p_l^{+,(k,l+1)} + p_l^{-,(k,l+1)}}{2} + f(t_l, q, i)\right\} \quad (50)$$

The minimization in (50) is carried out with a Newton's method, which in our problem is convergent because the quantity inside "{ }" in (50) is convex with respect to $q$. In most cases, the Newton method for our problem converges within ten cycles to achieve the absolute difference between the old and updated $q^*$ to be smaller than the sufficiently small value $10^{-12}$.

The discretization presented above is modified near the boundaries. For "$k=1$", set $p_l^{-,(k,l+1)} = \frac{\Phi_i^{(k,l+1)} - \Phi_i^{(k-1,l+1)}}{\Delta x}$. For $k = K-1$, set $p_l^{-,(k,l+1)} = \frac{\Phi_i^{(k+1,l+1)} - \Phi_i^{(k,l+1)}}{\Delta x}$. For $k = 0$, set $D_i^{(k,l)} = 0$ and $p_l^{+,(k,l+1)} = p_l^{-,(k,l+1)} = \frac{\Phi_i^{(k+1,l+1)} - \Phi_i^{(k,l+1)}}{\Delta x}$. For $k = K$, set $D_i^{(k,l)} = 0$ and $p_l^{+,(k,l+1)} = p_l^{-,(k,l+1)} = \frac{\Phi_i^{(k,l+1)} - \Phi_i^{(k-1,l+1)}}{\Delta x}$. Starting from the terminal value $\Phi_i^{(k,L)} = 0$ ($i \in J$, $0 \le k \le K$), the numerical solution $\Phi_i^{(k,l)}$ is recursively obtained backward in time at each vertex. Although not presented in this paper, we have checked that using a classical one-sided second-order upwind scheme at the boundaries does not affect the computational results presented below.

We employ the third-order accurate spatial discretization and combine it with the standard explicit Euler method. Then, under the assumption $\Delta t = O(\Delta v)$ commonly used in the explicit numerical methods, the resulting numerical scheme is only first-order accurate in both space and time. In this sense, it seems that using a higher-order temporal integration method is necessary. However, in the next sub-section, we show that numerical solutions are actually only first-order accurate in the space.

## 4.2 Test case

Computational performance of the presented numerical scheme is checked through its application to a test case. The model parameters are set as follows, so that the exact steady viscosity solution (23) exists: $\bar{V} = 1$, $(a,b) = (0.3, 0.7)$, $m = n = 1$, $y = 0.5$, $\underline{q} = 0$, $\bar{q} = 3$, $\tilde{q} = \ddot{q} = 1$, $w = 0.4$, $\delta = 0.1$, $\varpi \equiv 0$, and $I = 20$. The steady solution is approximated with the scheme by integrating the optimality equation (12) from $t = T = 125$ to $t = 0$ at which numerical solutions are found to be sufficiently close to a steady state. For given $\Delta v$, we set $0.25\Delta t$. Since the present test case is free from the regime switching we use aggregated switching rates of the realistic case ($I = 40$) identified in Section 4.3.

**Table 1** shows the $l^1$ and $l^\infty$ errors between the exact and numerical solutions. The results of the local Lax-Friedrichs scheme without the WENO reconstruction are also presented. **Figure 3** compares the exact and numerical solutions, graphically demonstrating that the scheme with the WENO reconstruction more accurately capture the sharp profile of the exact solution. We see that employing the WENO reconstruction is not always globally superior over the domain. This is considered because the smoothness indicator $r$ determining the effective stencils is not optimal for the present problem, but its optimization is beyond the scope of this paper.

The computational results suggest first-order convergence of the presented finite difference scheme. In addition, we see that the WENO reconstruction certainly improves the computational accuracy of the original local Lax-Friedrichs scheme especially near the points where the exact solution is non-smooth. The first-order accuracy of the scheme, despite the formal third-order accuracy of the WENO, is considered due to the non-smoothness of the exact solution (12). Nevertheless, we can see that using the WENO reconstruction can significantly improve computational accuracy of the scheme.

*Remark 5*

Our computational results empirically demonstrate convergence of numerical solutions generated by non-monotone schemes toward a non-smooth viscosity solution.

**Table 1.** Computed $l^1$ and $l^\infty$ errors between the exact and numerical solutions. LLxF: The local Lax-Friedrichs scheme without the WENO reconstruction. WENO: The scheme with the WENO reconstruction.

| | $l^1$ error | | $l^\infty$ error | |
|---|---|---|---|---|
| K | LLxF | WENO | LLxF | WENO |
| 50 | 0.00771 | 0.00659 | 0.01980 | 0.01202 |
| 100 | 0.00364 | 0.00323 | 0.01005 | 0.00599 |
| 200 | 0.00176 | 0.00160 | 0.00507 | 0.00299 |
| 400 | 0.00087 | 0.00080 | 0.00255 | 0.00149 |
| 800 | 0.00043 | 0.00040 | 0.00128 | 0.00075 |

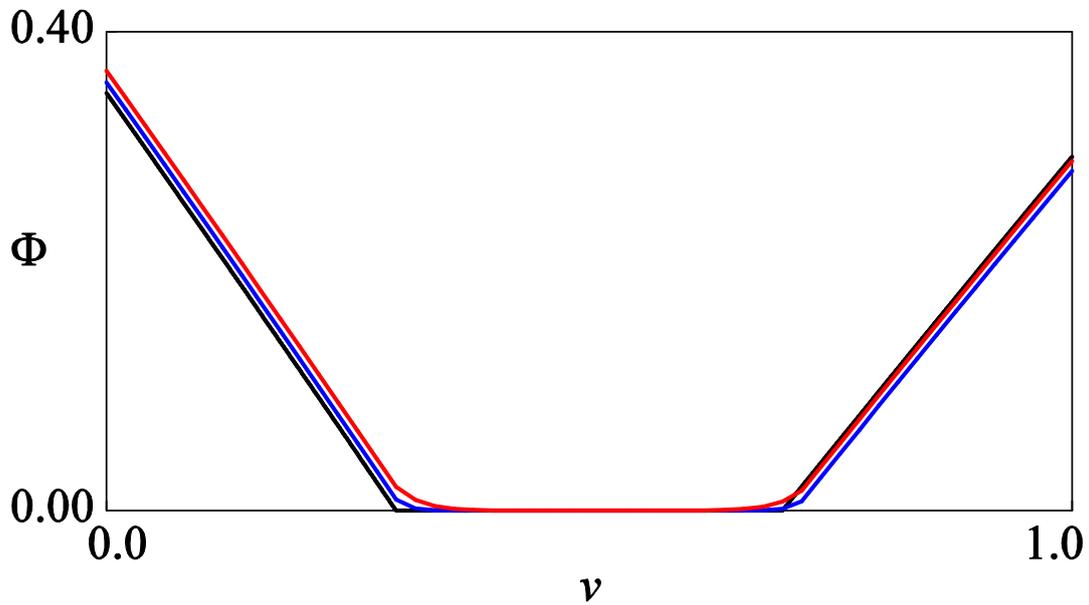

**Figure 3.** Comparison of the exact and computed value functions ($K = 50$). Black: exact solution, Red: original local Lax-Friedrichs scheme, and Blue: the scheme with the WENO reconstruction. The lateral axis is the water volume and the vertical axis represents the exact and numerical solutions.

## 4.3 Application

### 4.3.1 Parameter estimation

The model parameters have been estimated for Obara Dam installed at a middle reach of Hii River, Japan. The dam has a reservoir with the capacity of $\bar{V} = 6.08 \times 10^7$ (m³) and has been operated from 2011 for water resources supply and flood mitigation [54]. The maximum outflow discharge is designed to be larger than $\bar{q} = 250$ (m³/s) and the minimum base outflow discharge is $\underline{q} = 1$ (m³/s). This dam has been chosen in the model application because public hourly data of the inflow is available [54], and the dam-downstream environment has recently been a concern for local governments and fishery cooperatives [5, 38, 56]. Especially, the bloom of nuisance benthic algae *Cladophora glomerata* due to low outflow discharge has been a serious ecological concern. Our analysis therefore focuses on relatively low outflow discharges.

The record period of the hourly data utilized is from April 1 in 2016 to September 31 in 2019. The total number of the data is 31,417. The available data are categorized into the following $I+1 = 41$ regimes: $R_i = [\Delta_i, \Delta_{i+1})$ with $\Delta_i = 10i$ ($0 \le i \le 40$) (m³/s) and $\Delta_{41} = +\infty$ (m³/s). In addition, the representative values of the discharges for each of the regimes are set as $Q_i = 5.0i + 2.5$ (m³/s). The basic operation policy of the dam is designed to be $\ddot{q} = Q_i$ [55]. By the field surveys in the downstream reach of Obara Dam, the authors have found that there is almost no algae bloom when the outflow discharge is larger than about 15 (m³/s), suggesting to set $\tilde{q} = 15$ (m³/s). Therefore, if $q^*$ is close to $\ddot{q} = Q_i$, then the algae bloom can be effectively suppressed when $i \ge 1$.

The hourly switching probabilities $p_{ij} = \Pr(\theta_{t+h} = Q_i | \theta_t = Q_j)$ with $h = 1$ (h) is estimated from the available data. The quantity $p_{ij}$ represents the probability of switching the regime from $i$ to $j$ during the time interval $(t, t+h)$. We assume that $p_{ij}$ is time-homogenous for the sake of simplicity of analysis. Using the estimated $p_{ij}$, the switching rates $\lambda_{ij}$ (1/h) are estimated as

$$\lambda_{ij} = \begin{cases} \dfrac{1 - p_{ij}}{h} & (i = j) \\ \dfrac{p_{ij}}{h} & (i \neq j) \end{cases}. \tag{51}$$

The switching probabilities $p_{ij}$ are empirically estimated as in **Figure 4**. The estimation results imply a nearly diagonal structure of the matrix $\{p_{ij}\}_{0 \le j \le I}$ especially for relatively low flows, meaning that the low inflow regimes are more persistent than the higher ones. In fact, the corresponding stationary probability density of the estimated Markov chain equal 0.846, 0.0486, 0.0198, and 0.0131 for the regimes $i = 0,1,2,3$, respectively. **Figure 4** implies that the jumps such that $|i-j|$ is large are rare because of its diagonal-like structure.

The other model parameters have been set as follows: $(a,b) = (0.2\bar{V}, 0.8\bar{V})$, $m = n = 1$, $y = 0.5 \times \breve{q}^2$ with $\breve{q} = 10$ (m³/s) (to consistently set the unit of each term in the performance index), $w = 0.4$, and $\delta = 0.02$ (1/day). For the sake of simplicity of analysis, we set $\varpi \equiv 0$, assuming that the inflow and outflow discharges dominate the water balance dynamics.

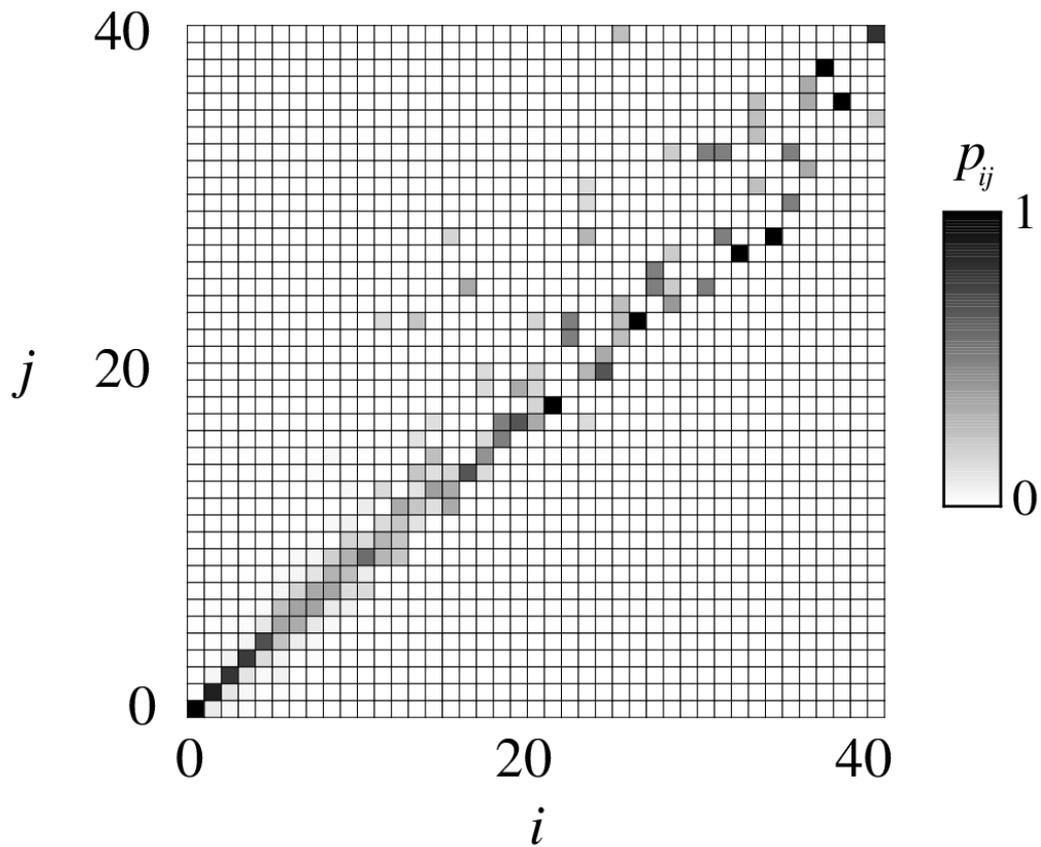

**Figure 4.** Estimated switching probabilities $p_{ij}$ from the regimes $i$ to $j$.

### 4.3.2 Computational results

All the numerical computations below have been carried out with the resolution with $T = 500$ (day), $L = 200,000$, $K = 400$, and thus $\Delta t = 0.0025$ (day) and $\Delta v = 0.0025\overline{V}$. Starting from the backward temporal integration from the terminal time $t = T$, the obtained numerical solutions are found to be sufficiently close to the steady state, in the sense of the $l^\infty$-norm between the numerical solutions at successive time steps, at the initial time $t = 0$. The temporal factor $86,400/\overline{V}$ and the volume factor $\overline{V}$ are used in the computation so that the dimension of the time becomes day and that of the volume becomes non-dimensional. The plotted value functions are normalized ones.

**Figures 5** and **6** show the computed value function $\Phi$ and the optimal outflow discharge $q^*$ for all the regimes ($0 \leq i \leq I = 40$). Being different from the previous exact solution, the numerical solution seems to be smooth. The computational results in the figures suggest that the optimal outflow discharge is monotonically increasing with respect to $i$ and the profiles are qualitatively close to those of the exact solution. On the computed value function, the profiles of $\Phi_i$ are monotonically decreasing with respect to $i$. Being different from the exact solutions derived in the previous section, the computed value functions do not vanish for the moderate range of the water volume because of the regime-dependent and thus non-constant $\breve{q} = Q_i$.

The inflow discharge $Q_i$ is assumed not to be greater than the threshold discharge $\breve{q}$ on the downstream algae bloom for $i = 0, 1, 2, 3$. The value functions $\Phi_i$ are decreasing with respect to $i$ in most parts of the domain because choosing simply $q = \breve{q} = Q_i$ activates the penalty (8) for relatively small $i$, but triggers no problem for larger $i$ unless the water volume is smaller than the threshold $v = a$. The computed optimal outflow discharges $q^*$ are state-dependent and increasing with respect to $i$ as in the exact solution derived in the previous section. A difference is that it is not monotone with respect to the water volume $v$ for each regime $i$. The unimodal profile of each $\Phi_i$ for the relatively small water volume $0 < v < a$ is due to the conflicting objective that the water volume should be increased to $a \leq v \leq b$, while the outflow discharge should be sufficiently large to inactivate the penalty (8) as possible.

Another computational example is also presented for a more complicated case where too large outflow discharges are also penalized. We then specify the coefficient $f$ as $f = f_1 + f_2 + f_3$, where

$$f_3(q_s) = \frac{y}{m+1}\max\{q_s - q', 0\}^{m+1} \tag{52}$$

with another threshold discharge $q' = 50$ (m³/s), above which the new penalty function (52) is activated. For example, too large discharge may flush out aquatic species toward downstream reaches [57, 58], increasing ecological disutility.

**Figures 7** and **8** show the computed value function $\Phi$ and the optimal outflow discharge $q^*$ for all the regimes ($0 \leq i \leq I = 40$) with the augmented $f$. The value functions and the optimal outflow discharge are qualitatively different from the previous ones. Especially, the value functions are significantly different among the regimes are larger than the previous ones. The latter fact is mainly due to using a larger $f$ having an additional term in this case. In the present case, the optimal outflow discharge is smaller than the previous one due to penalizing both small and large outflow discharges. This tendency is more clearly seen in the inflow regimes corresponding to relatively high inflows. The decision-maker with the new performance index encounters a more complicated operation of the dam-reservoir system. Nevertheless, the presented computational results fully characterize his/her optimal strategy as a function of the inflow regime $i$ and the water volume $v$ at each time.

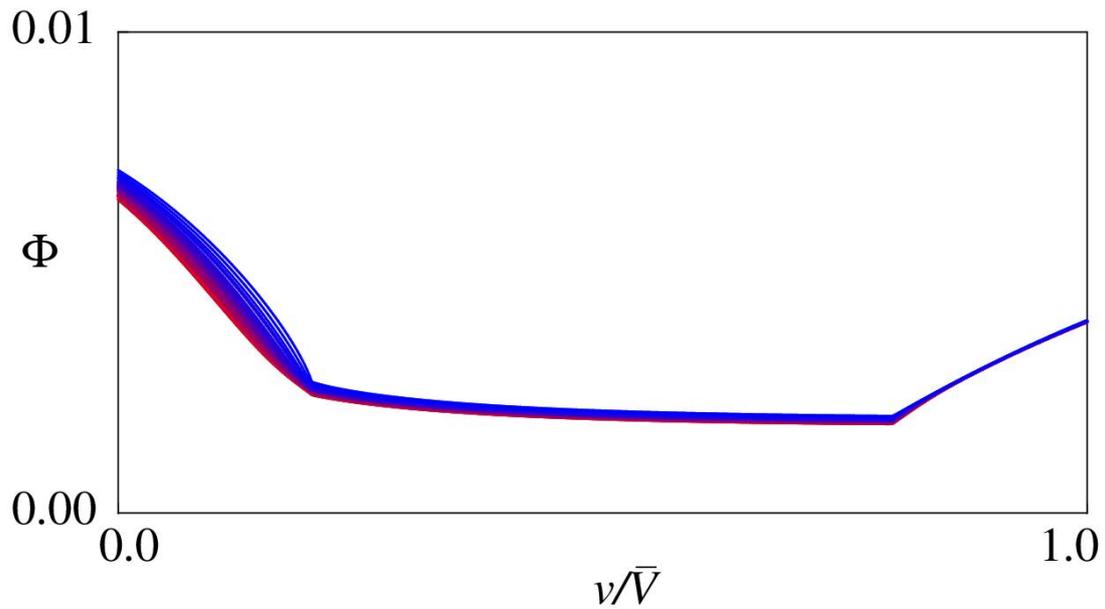

**Figure 5.** Computed value functions $\Phi_i$ for all the regimes $0 \leq i \leq I = 40$. The index $i$ increases as the color becomes from the blue toward the red. The lateral axis is the normalized water volume and the vertical axis represents the value functions.

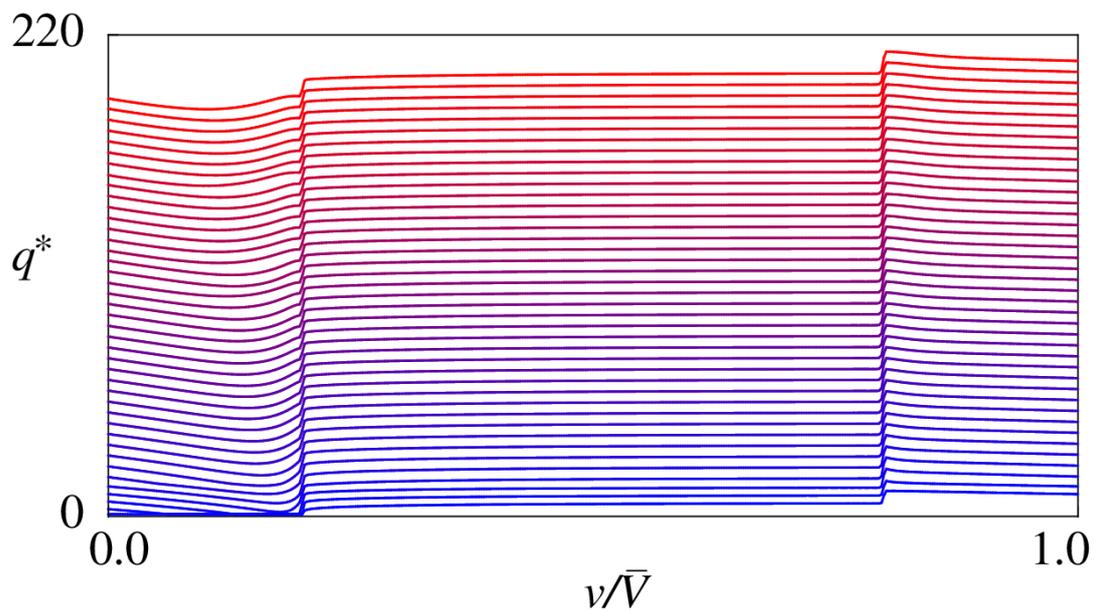

**Figure 6.** Computed optimal outflow discharge $q_i^*$ (m$^3$/s) for all the regimes $0 \leq i \leq I = 40$. The index $i$ increases as the color becomes from the blue toward the red. The lateral axis is the normalized water volume and the vertical axis represents the optimal controls.

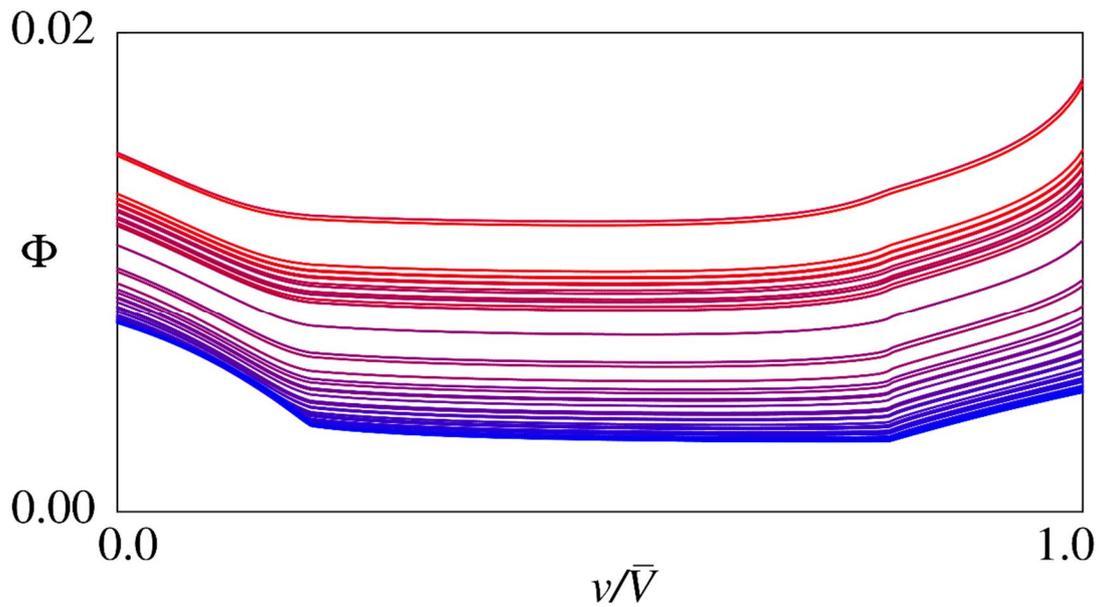

**Figure 7.** Computed value functions $\Phi_i$ for all the regimes ($0 \leq i \leq I = 40$) with the augmented $f$. The index $i$ increases as the color becomes from the blue toward the red. The lateral axis is the normalized water volume and the vertical axis represents the value functions.

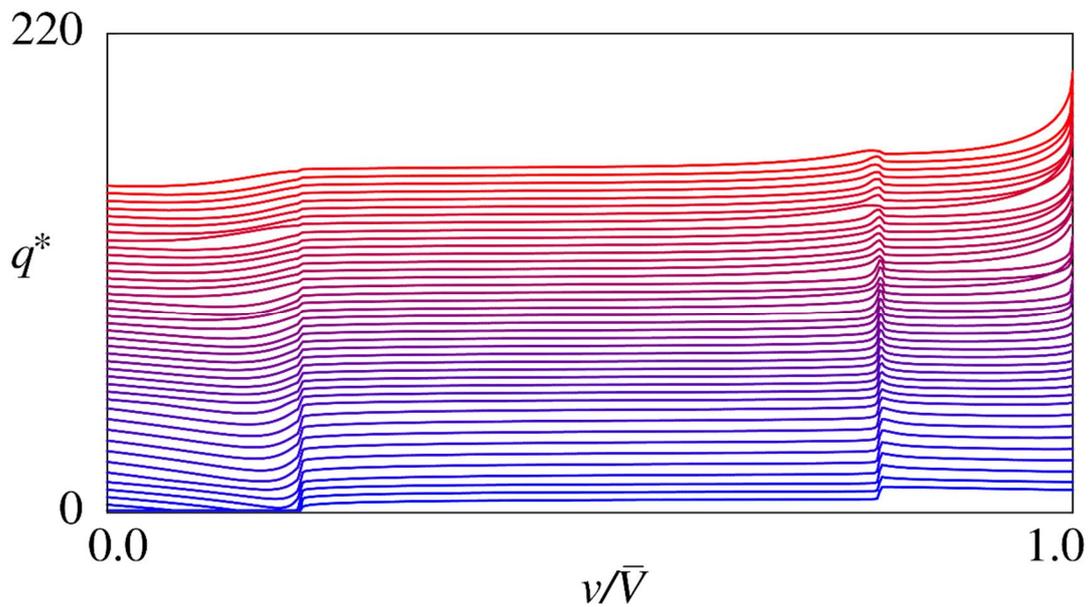

**Figure 8.** Computed optimal outflow discharge $q_i^*$ (m$^3$/s) for all the regimes ($0 \leq i \leq I = 40$) with the augmented $f$. The index $i$ increases as the color becomes from the blue toward the red. The lateral axis is the normalized water volume and the vertical axis represents the optimal controls.

## 5. Conclusions

We formulated a stochastic control problem of a dam-reservoir system created in a river. The regime-switching description of the system dynamics harmonized with the dynamic programming principle and effectively reduced the optimization problem to a terminal problem of the optimality equation. Solutions to the optimality equation were characterized from the viewpoint of constrained viscosity solutions. An exact steady viscosity solution was found under a simplified condition. The solution is continuous but non-smooth, and is therefore not a classical solution satisfying the equation point-wise. A local Lax-Friedrichs scheme equipped with a WENO reconstruction was presented and verified against the derived non-smooth exact solution. The model was finally applied to a problem of Obara Dam created in Hii River, Japan with identified model parameter values. The application results implied how the operation policy should be adapted according to the environmental concerns.

There is a variety of ways to extend the presented mathematical model. The inflow discharge would follow climate changes in future, and the Markov chain can be updated adaptively based on future hydrological information. Statistically learning the inflow time-series using a probabilistic forecast method [59] or using a reinforcement learning method [60] is an option to customize the presented model. The presented model can work as an upstream boundary condition of the eco-hydraulic model [61]. We must pay attention to multiple local optima [62] in these cases. Adding other variables to the model, such as population dynamics of aquatic species [38, 63] inside and/or downstream of a reservoir is also an interesting topic. Adding a reliability constraint [64] to the model enables us to derive optimal controls guaranteeing certain reliability criteria. Considering disaster management based through operating dams is an important engineering problem as well [65], which can be addressed by revising the performance index of the presented model. Model ambiguity, which is due to uncertainties of the coefficients and parameter values, can be severe in data-sparse cases [66]. This issue can be addressed through the utilization of the modern robust control framework, such as the nonlinear expectation [67] and some entropic penalization techniques [68, 69]. Boundary treatment of the optimality equation has to be modified if some control strategy at the extremes should be prescribed *a priori* [70]. Finally, finding a numerical method to achieve truly higher-order accuracy of WENO schemes for the optimality equation having the discontinuous source term is an important topic from both theoretical and practical viewpoints because it would realize a more efficient computation.

**Acknowledgements**

JSPS Research Grant No. 19H03073, Kurita Water and Environment Foundation Grant No. 19B018, and a grant for ecological survey of a life history of the landlocked ayu *Plecoglossus altivelis altivelis* from the Ministry of Land, Infrastructure, Transport and Tourism of Japan support this research. This research was carried out under a support of the research fund for young researchers in Shimane University. The authors thank the anonymous reviewer for his/her careful reading and helpful suggestions and comments